\documentclass[aps,pra,reprint,showpacs,amsmath,amssymb]{revtex4-1}
\usepackage{}

\usepackage{graphicx}
\usepackage{bm}
\usepackage{amssymb}
\usepackage{amsmath}
\usepackage{amsthm}
\usepackage{amsfonts}
\usepackage{mathrsfs}
\usepackage{hyperref}

\begin{document}

\title{Third-order GHZ-type and W-type Spatial Correlations with Classical Lights}

\author{Er-Feng Zhang}
\author{Quan Li}
\author{Wei-Tao Liu}\email{wtliu@nudt.edu.cn}
\author{Hui-Zu Lin}
\author{Ping-Xing Chen}\email{pxchen@nudt.edu.cn}
\affiliation{College of Science, National University of Defense Technology, Changsha 410073, China}
\date{\today}
\begin{abstract}
The spatial correlation with classical lights, which has some similar aspects as that with entangled lights, is an interesting and fundamentally important topic. But the features of high-order spatial correlation with classical lights are not well known, and the types of high-order correlations produced are of limit. Here, we propose a scheme to produce third-order spatial correlated states by modulating the phases of three laser beams. With the scheme we can produce Greenberger-Horne-Zeilinger-type (GHZ-type) and W-type spatial correlations with different phase modulations. Our scheme can be easily generalized to produce $N$-order spatial correlation states and to probe the aspects of different multi-partite spatial correlations.
\end{abstract}

\maketitle

The high-order spatial correlation effect has been of great importance in optics \cite{1}. The investigations on this effect not only can help us to understand the nature of light more effectively \cite{3}, but also make it advantageous to bring such effect into practice \cite{5,6,7,8,9,10,11,12,13,14,15,16}.

In 1995, second-order spatial correlation phenomenon was observed experimentally with entangled photon pairs from spontaneous parametric down-conversion (SPDC) \cite{5,6}, which led to the researches about the theories and applications of the second-order spatial correlation effect with entanglement \cite{7,8,9}. Several years later, it was discovered that many of the features of the second-order spatial correlation obtained with entangled photon pairs, such as ghost imaging, subwavelength interference and so on, can also be realized with classical lights \cite{10,11,12,13,14,15,16}. Then, the third-order spatial correlation effect has been extensively studied with both entangled and classical lights \cite{36,37,24,25,38,26,27,28,29,30}. But the features and types of high-order spatial correlation are not well known to us for both entangled and classical lights. As we know, for entangled lights, there are GHZ and W entangled states in three-partite system and more complex types of entangled states in multi-partite system \cite{32,31}. The different types of entangled states can produce corresponding spatial correlations \cite{36,37,38}. For classical lights, we do not know how to produce GHZ-type and W-type correlations in the three-partite system, and do not know the types of correlation in general multi-partite systems.

In this work, we focus on high-order spacial correlation with classical lights. First we define the GHZ-type correlated state and W-type correlated state. Then, we present a scheme to produce GHZ-type and W-type correlated states with three classical light sources generated by applying three sets of time-variable phase masks onto three laser beams, respectively. It is shown that when the three sets of phase masks are different but correlated, the state of the three classical lights is a GHZ-type spatial correlated state. When the three sets of phase masks are identical, the state of the three classical lights is a W-type spatial correlated state. This scheme can be generalized to multi-partite system, which means that we can produce multi-partite correlated states, and achieve different higher-order spatial correlations.

We define the GHZ-type correlated state as a state with which one can achieve third-order correlation, but no second-order correlations, and the W-type correlated state with which one can achieve both third-order and second-order correlations. It should be noted that the definition of GHZ-type (W-type) correlated states are just similar to the GHZ-type (W-type)entangled states in which there are three-partite entanglement but no (have) two-partite entanglement. Furthermore, for multi-partite systems the types of correlations are not known as the entanglement does. Although the features of correlated states may be simpler than those of the entangled states, they are still unclear to us and the correlated states have many theoretical and experimental applications, such as non-locality without entanglement and correlation imaging.

\begin{figure}
\begin{center}
\includegraphics [width=\columnwidth]{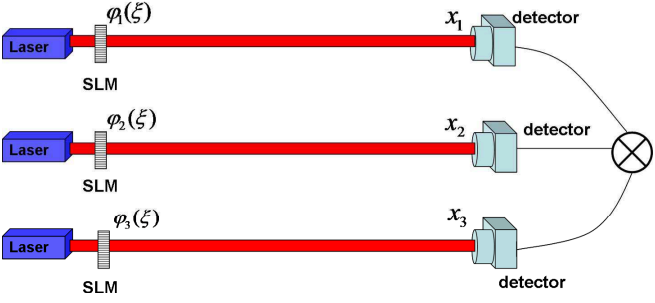}
\end{center}
\caption{\label{fig:1}(color online). Standard schematic of third-order spatial correlation using classical lights. Laser: laser light source; SLM: spatial light modulator; detector: the detector to measure the field pattern of the light source.}
\end{figure}

With the definition above, we present a scheme to produce high-order correlation. A schematic diagram of the model for describing the third-order spatial correlation effect is shown in Fig. \ref{fig:1}. Three classical light sources are involved. Each classical source is obtained by applying a set of phase masks onto a laser beam. These phase masks can be achieved with a spatial light modulator (SLM). The values in these phase masks $\varphi_n(\xi, k)$ for the $n$-th $(n=1,2,3)$ classical source are spatially random-distributed and time-varying, with $\xi$ being the position vector on the phase masks, and $k=1, \cdots, K$ being the index of the phase modulation samples. Here, we consider that the three light sources are placed at one same coordinate system. It will be shown that we can obtain GHZ-type and W-type spatial correlations among the three classical sources by properly controlling the three phase masks.

Here, we consider the single photon model of each laser beam. Then, the state of the laser beams can be written as
\begin{equation}\label{eq:laser}
|\psi^{(n)}\rangle=\int d\xi |1^{(n)}(\xi)\rangle, (n=1,2,3),
\end{equation}
where $|1_{\xi}^{(n)}\rangle$ presents the state that one photon is emitted from the position $\xi$ in the $n$-th laser beam. The state in Eq. \eqref{eq:laser} means that the single-photon state in every position $\xi$ is coherent. After the phase modulations, the three-photon density matrix to describe the three light sources can be written as \cite{33,34}
\begin{equation}\label{eq:1}
\begin{split}
\rho_3=&\int d\xi_1 d\xi_2 d\xi_3 d\xi_4 d\xi_5 d\xi_6 \left|1_{\xi_1}^{(1)}\right\rangle\left|1_{\xi_2}^{(2)}\right\rangle\left|1_{\xi_3}^{(3)}\right\rangle\\
&\cdot \left\langle 1_{\xi_4}^{(3)}\right|\left\langle 1_{\xi_5}^{(2)}\right|\left\langle 1_{\xi_6}^{(1)}\right|\cdot \langle \exp\{i[\varphi_1(\xi_1, k)+\varphi_2(\xi_2, k)\\
&+\varphi_3(\xi_3, k)-\varphi_3(\xi_4, k)-\varphi_2(\xi_5, k)-\varphi_1(\xi_6, k)]\}\rangle,
\end{split}
\end{equation}
where $\langle\cdot\rangle=\frac{1}{K}\sum_{k=1}^K\cdot$ is defined to be the sample average over the $K$ phase modulations. To investigate the second-order spatial correlation, two-photon state should be considered, which can be obtained by tracing away one photon of the three-photon state. Without loss of generality, the third photon is traced out, and the two-photon density matrix describing the other two light sources is \cite{33,34}
\begin{equation}\label{eq:2}
\rho_2=\int d\xi\left\langle1^{(3)}_{\xi}\right|\rho_3\left|1^{(3)}_{\xi}\right\rangle.
\end{equation}
Furthermore, the fields at any distances $d_n$ from the three light sources can be obtained using the Fresnel diffraction integral, respectively \cite{34,35}:
\begin{equation}\label{eq:3}
E_n^{(+)}(x_n)=\int d\xi a_n(\xi)h_n(x_n, \xi),
\end{equation}
where $a_n(\xi)$ is the annihilation operator in the $n$-th optical source. Moreover, $h_n(x_n, \xi)$ is the Green's function associated to the propagation of the field from the source to the detection plane \cite{35}:
\begin{equation}\label{eq:4}
h_n(x_n, \xi)\propto \exp\left[i\pi (x_n-\xi)^2/\lambda d_n\right],
\end{equation}
where $\lambda$ is the wavelength of the laser light source, and $x_n$ is the position vector at the detection plane. Then, by cross-correlating the field patterns, we can obtain the third-order and second-order spatial correlation functions \cite{3}:
\begin{equation}\label{eq:5}
\begin{split}
G_3(x_1, x_2, x_3)=&\text{Tr}[\rho_3 E_1^{(-)}(x_1)E_2^{(-)}(x_2)E_3^{(-)}(x_3)\\
&\times E_3^{(+)}(x_3)E_2^{(+)}(x_2)E_1^{(+)}(x_1)],\\
G_2(x_1, x_2)=&\text{Tr}[\rho_2 E_1^{(-)}(x_1)E_2^{(-)}(x_2)\\
&\times E_2^{(+)}(x_2)E_1^{(+)}(x_1)].
\end{split}
\end{equation}

\begin{figure}
\begin{center}
\includegraphics [width=\columnwidth]{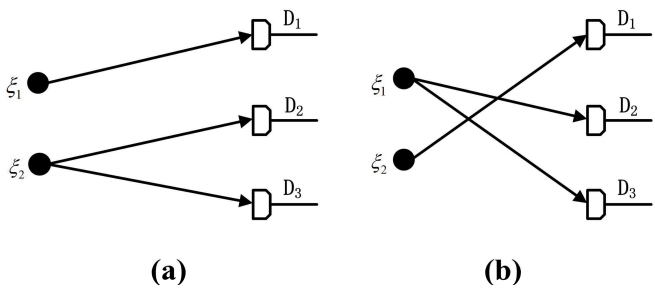}
\end{center}
\caption{\label{fig:2}The representation of the three-photon interference. It is the result of the superposition between these two indistinguishable three-photon amplitudes: (a), $h_1(x_1, \xi_1)h_2(x_2, \xi_2)h_3(x_3, \xi_2)$: the probability amplitude that a photon in position $\xi_1$ of field 1 goes to detector $\text{D}_1$, a photon in position $\xi_2$ of field 2 and 3 goes to detector $\text{D}_2$ and $\text{D}_3$, respectively; (b), $h_1(x_1, \xi_2)h_2(x_2, \xi_1)h_3(x_3, \xi_1)$: the probability amplitude that a photon in position $\xi_2$ of field 1 goes to detector $\text{D}_1$, a photon in position $\xi_1$ of field 2 and 3 goes to detector $\text{D}_2$ and $\text{D}_3$, respectively.}
\end{figure}

In this section, we will consider the situation of third-order GHZ-type spatial correlation. Here, the modulation method is considered: the values in the phase masks $\varphi_2(\xi, k)$ and $\varphi_3(\xi, k)$ are statistically independent, uniformly and randomly distributed in the range $[0, 2\pi)$, while keeping $\varphi_1(\xi, k)=\varphi_2(\xi, k)+\varphi_3(\xi, k)$. That is, the three sets of phase masks loaded on the three laser beams are different but correlated. In this case, we get obviously
\begin{equation}\label{eq:6}
\begin{split}
&\langle \exp[i\varphi_n(\xi, k)]\rangle=0, (n=1,2,3),\\
&\langle \exp\{i[\varphi_n(\xi, k)-\varphi_{n'}(\xi', k)]\}\rangle=\delta_{n,n'}\delta(\xi-\xi'), \\
&\langle \exp\{i[\varphi_1(\xi_1, k)-\varphi_2(\xi_2, k)-\varphi_3(\xi_3, k)]\}\rangle\\
&=\delta(\xi_1-\xi_2)\delta(\xi_1-\xi_3).
\end{split}
\end{equation}
In this way, the density matrix of the three-photon state can be written as
\begin{equation}\label{eq:7}
\rho_3\propto\int d\xi_1 d\xi_2d\xi_3|\psi_3(\xi_1,\xi_2,\xi_3)\rangle\langle\psi_3(\xi_1,\xi_2,\xi_3)|.
\end{equation}
Here, in the case that $\xi_1\neq\xi_2=\xi_3$, the term $|\psi_3(\xi_1,\xi_2,\xi_3)\rangle$ can be denoted by $|\psi_3(\xi_1,\xi_2)\rangle$, which is a GHZ-type state \cite{32}:
\begin{equation}\label{eq:8}
\begin{split}
|\psi_3(\xi_1,\xi_2)\rangle=&\frac{1}{\sqrt{2}}\left[\left|1^{(1)}_{\xi_1}\right\rangle\left|1^{(2)}_{\xi_2}\right\rangle
\left|1^{(3)}_{\xi_2}\right\rangle\right.\\
&\left.+\left|1^{(1)}_{\xi_2}\right\rangle\left|1^{(2)}_{\xi_1}\right\rangle\left|1^{(3)}_{\xi_1}\right\rangle\right].
\end{split}
\end{equation}
In the other cases, $|\psi_3(\xi_1,\xi_2,\xi_3)\rangle=\left|1_{\xi_1}^{(1)}\right\rangle\left|1_{\xi_2}^{(2)}\right\rangle\left|1_{\xi_3}^{(3)}\right\rangle$ is a product state. The GHZ-type state in Eq. \eqref{eq:8} results in the superposition of two different but indistinguishable alternatives: $h_1(x_1, \xi_1)h_2(x_2, \xi_2)h_3(x_3, \xi_2)$ and $h_1(x_1, \xi_2)h_2(x_2, \xi_1)h_3(x_3, \xi_1)$, the so-called three-photon interference, as shown in Fig. \ref{fig:2}. The third-order spatial correlation function corresponding to the GHZ-type state in Eq. \eqref{eq:8} is
\begin{equation}\label{eq:9}
\begin{split}
G_3(x_1, x_2, x_3)&\propto\int d\xi_1 d\xi_2 |h_1(x_1, \xi_1)h_2(x_2, \xi_2)h_3(x_3, \xi_2)\\
&  \quad +h_1(x_1, \xi_2)h_2(x_2, \xi_1)h_3(x_3, \xi_1)|^2 .
\end{split}
\end{equation}

When the distances $d_1$, $d_2$ and $d_3$ satisfy the condition:
\begin{equation}\label{eq:10}
\frac{1}{d_1}-\frac{1}{d_2}-\frac{1}{d_3}=0,
\end{equation}
Eq. \eqref{eq:9} can be simplified as
\begin{equation}\label{eq:11}
G_3(x_1, x_2, x_3)\propto 1+\frac{\sin^2\left[\frac{\pi D}{\lambda}\left(\frac{x_1}{d_1}-\frac{x_2}{d_2}-\frac{x_3}{d_3}\right)\right]}{\left[\frac{\pi D}{\lambda}\left(\frac{x_1}{d_1}-\frac{x_2}{d_2}-\frac{x_3}{d_3}\right)\right]^2},
\end{equation}
where $D$ is the size of the SLM. The term in Eq. \eqref{eq:11} means a perfect spatial correlation among the light at the three detection planes $(x_1$, $x_2$ and $x_3)$, which has possible applications, such as ghost imaging.

By tracing out the third photon in the three-photon density matrix in Eq. \eqref{eq:7}, the two-photon density matrix describing two of the three light sources is obtained:
\begin{equation}\label{eq:12}
\begin{split}
\rho_2&=\int d\xi\left\langle1^{(3)}_{\xi}\right|\rho_3\left|1^{(3)}_{\xi}\right\rangle\\
&=\prod_{n=1}^2\int d\xi_n \left|1^{(n)}_{\xi_n}\right\rangle\left\langle1^{(n)}_{\xi_n}\right|.
\end{split}
\end{equation}
This means that the two-photon density matrix is a product of the two single-photon density matrices $\int d\xi_n \left|1^{(n)}_{\xi_n}\right\rangle\left\langle1^{(n)}_{\xi_n}\right|$, $(n=1,2)$. As a result, there is no second-order spatial correlation.

\begin{figure}
\begin{center}
\includegraphics [width=\columnwidth]{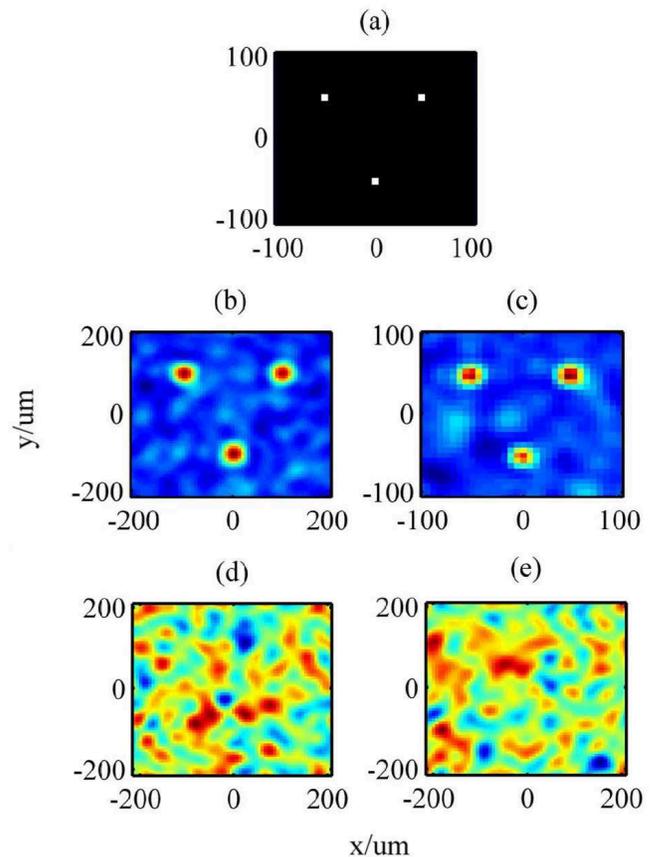}
\end{center}
\caption{\label{fig:3}(color online). Simulated results of third-order spatial correlation effect with an object of three points. (a) is the object to be imaged; (b) and (c) are the images reconstructed by measuring the third-order spatial correlation functions with the two types of measurement operations, respectively; (d) and (e) are the results obtained by measuring the second-order spatial correlation functions, respectively.}
\end{figure}

In order to verify the third-order GHZ-type spatial correlation effect, we performed numerical simulations, and the results are shown in Fig. \ref{fig:3}. As the input for simulations, we considered the following case. Three laser light sources with wavelength $\lambda=532$ $\text{nm}$ illuminate three SLMs with size $D=2$ $\text{mm}$, respectively. Then, the beam from the first source transmits through a three-point object shown in Fig. \ref{fig:3}(a), and a single-pixel detector collects the light transmitted from the object. The intensities of the other two light beams are measured by two scanning detectors. The distances are set as $d_1=10$ $\text{cm}$, $d_2=d_3=20$ $\text{cm}$, which guarantees the spatial correlation at the three detection planes. From Fig. \ref{fig:3}(b) and Fig. \ref{fig:3}(c), it is found that the information of the object can be obtained by the third-order spatial correlation functions. In addition, there are two kinds of measurement operations to reconstruct the image of the object. The first way is to move one detector $(x_3)$ with the other detector $(x_2=0)$ fixed, and the magnification factor of the image is $d_3/d_1=2$ with the image shown in Fig. \ref{fig:3}(b). The second way is to move both detectors $(x_2=x_3)$ together, with the result shown in Fig. \ref{fig:3}(c), and the magnification factor of the image is 1. However, the information of the object can not be reconstructed by second-order correlation functions, as shown in Fig. \ref{fig:3}(d, e). The simulation results above show the GHZ-type spatial correlated state can be attained with the modulation method, which can achieve third-order GHZ-type spatial correlation.

Now we will consider the W-type spatial correlation case where the three sets of phase masks loaded on the three SLMs satisfy the condition: the three sets of phase masks are the same, namely $\varphi_1(\xi, k)=\varphi_2(\xi, k)=\varphi_3(\xi, k)\triangleq\varphi(\xi, k)$, and the values of the phases added on the pixels in the SLMs are statistically independent, uniformly distributed in the range $[0, 2\pi)$. This case is equivalent to the model of observing one type of third-order spatial correlation effect with classical light in Ref. \cite{26,27,30}. As these phase masks obey the relationships: $\langle \exp[i\varphi(\xi, k)]\rangle=0$ and $\langle \exp\{i[\varphi(\xi_1,k)-\varphi(\xi_2,k)]\}\rangle=\delta(\xi_1-\xi_2)$, the three-photon density matrix to describe the three light sources is simplified to
\begin{equation}\label{eq:13}
\rho_3\propto\int d\xi_1d\xi_2d\xi_3|\psi_3(\xi_1,\xi_2,\xi_3)\rangle
\langle\psi_3(\xi_1,\xi_2,\xi_3)|.
\end{equation}
Here, in the case that $\xi_1\neq\xi_2\neq\xi_3$, the term $|\psi_3(\xi_1,\xi_2,\xi_3)\rangle$ can be taken as a W-type state \cite{31}:
\begin{equation}\label{eq:14}
|\psi_3(\xi_1,\xi_2,\xi_3)\rangle=\frac{1}{\sqrt{6}}\sum_P\prod_{n=1}^3\left|1^{(n)}_{\xi_{P(n)}}\right\rangle,
\end{equation}
where the sum $\sum_P$ runs over the all $3!$ possible permutations $P$ of the set of integers $1,2,3$. In the case that two of the three parameters $\xi_1$, $\xi_2$ and $\xi_3$ are the same and different from the other one, such as $\xi_1\neq\xi_2=\xi_3$, the term $|\psi_3(\xi_1,\xi_2,\xi_2)\rangle$ can also be taken as a W-type state \cite{31}:
\begin{equation}\label{eq:15}
\begin{split}
|\psi_3(\xi_1,\xi_2,\xi_2)\rangle=&\frac{1}{\sqrt{3}}\left[\left|1^{(1)}_{\xi_1}\right\rangle\left|1^{(2)}_{\xi_2}\right\rangle
\left|1^{(3)}_{\xi_2}\right\rangle\right.\\
&+\left|1^{(1)}_{\xi_2}\right\rangle\left|1^{(2)}_{\xi_1}\right\rangle\left|1^{(3)}_{\xi_2}\right\rangle\\
&\left.+\left|1^{(1)}_{\xi_2}\right\rangle\left|1^{(2)}_{\xi_2}\right\rangle\left|1^{(3)}_{\xi_1}\right\rangle\right].
\end{split}
\end{equation}
In the other cases, the states $|\psi_3(\xi_1,\xi_2,\xi_3)\rangle=\left|1_{\xi_1}^{(1)}\right\rangle\left|1_{\xi_2}^{(2)}\right\rangle\left|1_{\xi_3}^{(3)}\right\rangle$ are product states. Therefore, the three-photon density matrix describing the three light sources is the incoherent superposition of a set of W-type and product states $|\psi_3(\xi_1,\xi_2,\xi_3)\rangle$, leading to third-order spatial correlation.

By tracing the third photon in the three-photon density matrix in Eq. \eqref{eq:13}, we obtain the two-photon density matrix:
\begin{equation}\label{eq:16}
\rho_2\propto\int d\xi_1 d\xi_2|\psi_2(\xi_1,\xi_2)\rangle\langle\psi_2(\xi_1,\xi_2)|,
\end{equation}
where the terms $|\psi_2(\xi_1,\xi_2)\rangle$ in Eq. \eqref{eq:16} are Bell-type or product states \cite{33}:
\begin{equation}\label{eq:17}
|\psi_2(\xi_1,\xi_2)\rangle=
\begin{cases}
\frac{1}{\sqrt{2}}\sum_P\prod_{n=1}^2\left|1^{(n)}_{\xi_{P(n)}}\right\rangle,&\xi_1\neq\xi_2,\\
\left|1_{\xi_1}^{(1)}\right\rangle\left|1_{\xi_2}^{(2)}\right\rangle,&\text{others}.
\end{cases}
\end{equation}
This means the second-order spatial correlation still exists in this case. Hence, the W-type spatial correlated state is produced, which can realize the W-type spatial correlation effect.

In the analytical results above, it shows that the third-order GHZ-type and W-type spatial correlations can be realized with three classical light sources. Without loss of generality, this scheme can be generalized to achieve $N$-order GHZ-type and W-type spatial correlations with $N$ classical light sources. If the $N$ sets of phase masks satisfy the condition: $\varphi_1(\xi, k)=\sum_{n=2}^N\varphi_n(\xi, k)$ one can achieve the $N$-order GHZ-type spatial correlation. If the $N$ sets of phase masks loaded on the $N$ laser beams are the same one can achieve the $N$-order W-type spatial correlation. Specially interesting, if the $N$ sets of phase masks satisfy other conditions, we may produce the $N$-order spatial correlation with the types more than GHZ-type and W-type spatial correlations. This means that we may produce the inequivalent multi-partite correlated states and different types of spatial correlations, and probe the features and the applications of multi-partite correlation with classical lights.

In conclusion, we propose a scheme to produce the high-order spatial correlation with classical light sources. We show that GHZ-type and W-type correlated states can be produced by applying three sets of time-variable phase masks onto three laser beams, respectively. When the three sets of phase masks are different but correlated, the state of the three classical lights is a GHZ-type spatial correlated state. When the three sets of phase masks are identical, the state of the three classical lights is a W-type spatial correlated state. This scheme can be easily generalized to multi-partite system to produce more types of multi-partite correlated states, and achieve more kinds of high-order spatial correlations besides the GHZ-type and W-type correlations. These results can deepen the basic understanding of the high-order spatial correlation effect effectively which is fundamentally important. Furthermore, it may lead to novel high-order correlation effects and applications.

This work is supported by the National Natural Science Foundation of China (Project Nos. 11004248, 11374368, 61405251 and 61201332).

%

\end{document}